\documentclass[onecollarge,natbib]{svjour2}
\bibpunct{[}{]}{;}{n}{}{,} 
\smartqed  
\usepackage{graphicx}
\journalname{Few-Body Systems (EFB22)}
\begin{document}

\title{Hadron structure within the point form of relativistic quantum mechanics
}
%

\author{M. G\'omez-Rocha \and W. Schweiger \and O. Senekowitsch
}


\institute{M. G\'omez-Rocha \and W. Schweiger \and O. Senekowitsch \at
              Institut f\"ur Physik, FB Theoretische Physik, Universit\"at Graz, Austria\\
              \email{wolfgang.schweiger@uni-graz.at}           
}

\date{Received: date / Accepted: date}

\maketitle

\begin{abstract}
We present an outline of recent developments in the field of
hadron form-factor  calculations within constituent-quark models
using the point form of relativistic quantum mechanics. Our method
to calculate currents and form factors is exemplified by means of
the weak $B\rightarrow D$ transition. We present results for weak
$B\rightarrow D$ transition form factors in the space- and the
time-like momentum-transfer region. We discuss how wrong cluster
properties, which one has to deal with when employing relativistic
quantum mechanics, affect these form factors and we estimate the
role non-valence, Z-graph contributions may play for decay
kinematics.

\end{abstract}

\vspace{-0.2cm}
\section{Introduction}
\label{intro}
In
Refs.~\cite{Biernat:2009my,Biernat:2010tp,Biernat:2011mp,GomezRocha:2012zd,Gomez-Rocha:2013bga}
the point form of relativistic quantum mechanics has been
advocated as an appropriate framework for calculating the
electroweak structure of bound few-body systems, in particular of
mesons and baryons within the scope of constituent-quark models.
The route to derive electroweak form factors, pursued in these
papers and followed also here, is to describe the physical process
in which particular form factors are measured in a Poincar\'e
invariant way, calculate the invariant 1$\gamma$- or 1$W$-exchange
amplitude, extract the hadronic current from this amplitude,
analyze its covariant structure and identify the form factors.
Poincar\'e invariance is thereby guaranteed by employing the
Bakamjian-Thomas construction~\cite{Bakamjian:1953kh}. The essence
of the Bakamjian-Thomas construction in point form is that the
4-momentum operator factorizes into an interaction-dependent mass
operator and a free 4-velocity operator
\begin{equation}\label{eq:massop}
\hat{P}^{\mu}=\hat{\mathcal M}\,  \hat V^{\mu}_{\mathrm{free}}=
\left(\hat{\mathcal M}_{\mathrm{free}}+ \hat{\mathcal
M}_{\mathrm{int}} \right) \hat V^{\mu}_{\mathrm{free}}\, .
\end{equation}
The dynamics of the system is thus completely encoded in the mass
operator.  From Eq.~(\ref{eq:massop}) it is quite obvious that all
four components of the 4-momentum operator are interaction
dependent. The generators of Lorentz transformations, on the other
hand, are interaction free. These are the properties which
characterize the point form of relativistic dynamics. They make it
comparably simple to boost and rotate wave functions and add
angular momenta.

Since $\gamma$- and $W$-exchange are treated dynamically, one has
to allow for particle emission and absorption. This is
accomplished by using a coupled-channel framework with a matrix
mass operator $\hat{\mathcal M}$ that acts on the direct sum of
the pertinent multiparticle Hilbert spaces. The diagonal entries
$\hat{M}_i$ of $\mathcal{M}$ are the sum of the relativistic
kinetic energies of the particles in channel $i$. In addition, the
$\hat M_i$ may contain instantaneous interactions between the
particles, like the confinement potential between quark and
antiquark. Off-diagonal entries of $\mathcal{M}$ are vertex
operators $\hat{K}_{i\rightarrow j}$ and $\hat{K}_{j\rightarrow
i}=\hat{K}_{i\rightarrow j}^\dag$ which describe the absorption
and emission of particles and hence the transition from one
channel to the other. A most convenient basis to represent all
these operators is formed by a complete set of velocity
states~\cite{Klink:1998zz}. A velocity state is a multiparticle
momentum state in the rest frame that is boosted to an overall
4-velocity $V$ ($V_\mu V^\mu=1$) by means of a rotationless boost
$B_c(V)$:
\begin{equation}\label{eq:vstates}
\vert V;  {\bf k}_1, \mu_1; {\bf k}_2, \mu_2; \dots ; {\bf k}_n,
\mu_n\rangle := \hat{U}_{B_c(V)} \vert {\bf k}_1, \mu_1; {\bf
k}_2, \mu_2; \ldots ; {\bf k}_n, \mu_n\rangle\quad \mathrm{with}
\quad \sum_{i=1}^{n}{{\bf k}_i}=0\, .
\end{equation}
The $\mu_i$s are the spin projections of the individual particles.
Using velocity states, matrix elements of vertex operators can be
simply related to field theoretical interaction Lagrangean
densities \cite{Klink:2000pp}
\begin{equation}
\label{vertexop} \langle V^{\prime}; { {\bf k}^{\prime}_{j},
\mu_{j}^{\prime} } \vert \hat{K} \vert V; {\bf k}_i, \mu_i \rangle
\propto V^0\, \delta^3({\bf V} - {\bf V}^{\prime})\, \langle {\bf
k}^{\prime}_j, \mu_j^{\prime} \vert
\hat{\mathcal{L}}_{\mathrm{int}}(0)\vert { {\bf k}_i, \mu_i }
\rangle \, .
\end{equation}
At this point it is worthwhile to remark that conservation of the
overall 3-velocity  at interaction vertices is a specific feature
of the Bakamjian-Thomas construction and does not hold, in
general, for point-form quantum-field theories. It is this overall
velocity-conserving delta function that leads to wrong cluster
properties, an unwanted feature of the Bakamjian-Thomas
construction which is observed in any form of relativistic
dynamics and is not just specific to the point
form~\cite{Keister:1991sb}. The physical consequences of wrong
cluster properties in our case are that the gauge-boson-hadron
vertices, which we analyze to obtain the transition form factors,
may not only depend on the momenta attached to the vertex, but
also on the lepton momenta. Formally such wrong cluster properties
could be cured by means of, so called, \lq\lq packing
operators\rq\rq, but practically these are hard to construct. The
strategy to obtain sensible results for the form factors  adopted in
Refs.~\cite{Biernat:2009my,Biernat:2010tp,Biernat:2011mp,GomezRocha:2012zd,Gomez-Rocha:2013bga}
and also followed here is thus rather to look for kinematical situations in which those spurious dependencies are minimized or vanish.

\vspace{-0.3cm}
\section{Weak $B\rightarrow D$ transition form factors}
\label{sec:ffs}
%
\subsection{Space-like momentum transfers}
\label{sec:ffspace}
\begin{figure}[t!]
\includegraphics[width=0.42\textwidth]{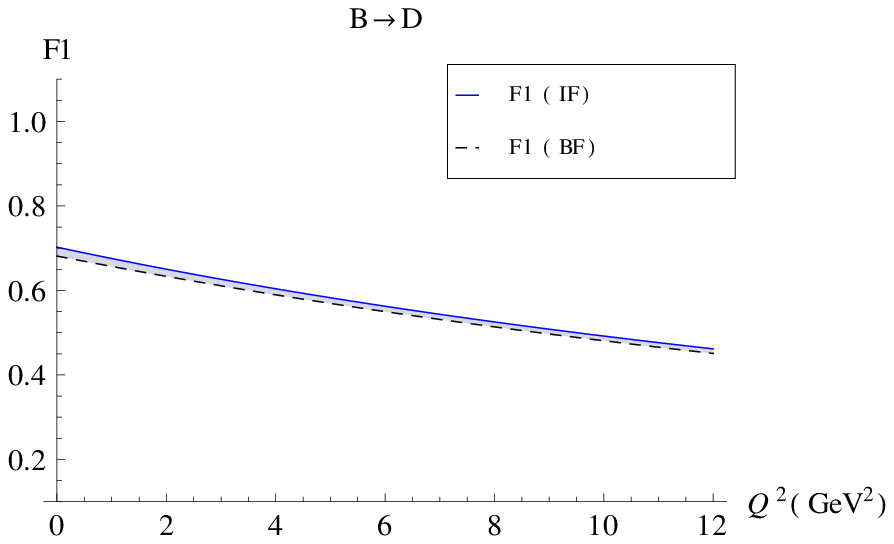}\hfill
\includegraphics[width=0.42\textwidth]{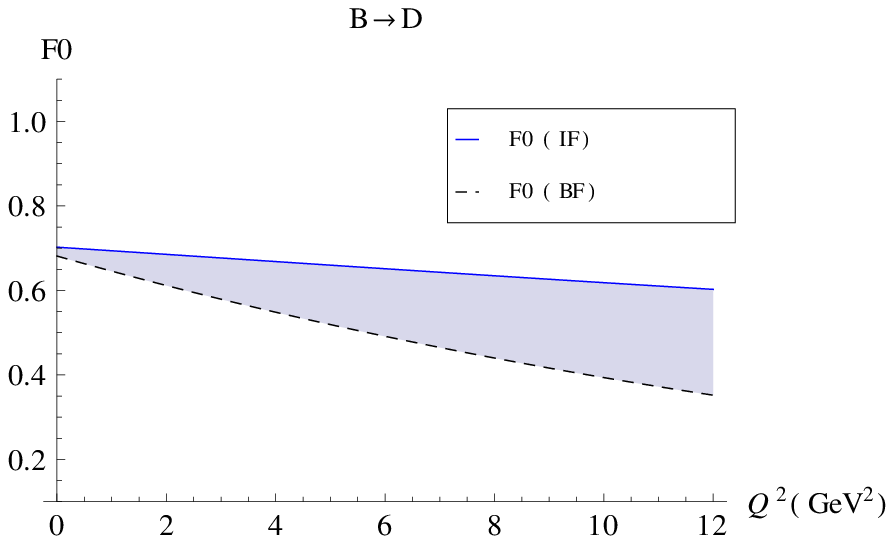}
\caption{The weak form factors $F1$ and $F0$ for the
$B\rightarrow D$ transition  in the space-like momentum-transfer
region as functions of $Q^2=-(p_B-p_D')^2$. The solid and dashed
lines refer to calculations in the infinite-momentum frame and the
Breit frame, respectively. The shaded area indicates the frame
dependence caused by the violation of cluster separability.}
\label{bdspacelike}
\end{figure}
Our first goal is to derive weak $B\rightarrow D$ transition form factors for space-like momentum transfers as they can, in principle, be measured in $\nu_e\, B^-\rightarrow e^- D^0$ scattering. In order to account for dynamical exchange of $W$-bosons we set up a 4-channel problem that includes all states which occur during such a scattering process if considered within a valence-quark picture (i.e. $|\nu_e, b, \bar u \rangle$, $|e, W^+, b, \bar u \rangle$, $|e, c, \bar u \rangle$, $|\nu_e, W^-, c, \bar u \rangle$). An instantaneos confinement potential between quark and antiquark is included in the diagonal entries of the matrix mass operator. Using perturbation theory for the weak coupling we  calculate the invariant  1$W$-exchange amplitude for $\nu_e\, B^-\rightarrow e^- D^0$ scattering. It is the sum of two time-ordered contributions which, as expected, is proportional to the contraction of a lepton with a meson current times the covariant $W$-boson propagator. This allows us to identify the weak meson current in a unique way. It is an overlap integral of $B$- and $D$-meson wave functions multiplied by the weak quark current, a Wigner-rotation factor and a kinematical factor (see, e.g., Refs.~\cite{GomezRocha:2012zd,Gomez-Rocha:2013bga}). After replacing the CM momenta $k_B$ and $k_D'$  by the respective physical particle momenta $p_B=B_c(V) k_B$ and $p_D'=B_c(V) k_D'$ we end up with a meson current that transforms like a 4-vector and has the covariant decomposition~\cite{GomezRocha:2012zd,Gomez-Rocha:2013bga} (with $q=p_B-p_D'$):
\begin{equation}
 J^{\mu}({\bf p}_D',{\bf p}_B)=\left[(p_B+p_D')-\frac{m_B^2-m_D^2}{q^2}\, q\right]^\mu\, F_1(Q^2,s)+
 \frac{m_B^2-m_D^2}{q^2}\, q^{\mu}\, F_0(Q^2,s)\, .
\end{equation}
The physical consequences of wrong cluster properties, inherent in the Bakamjian-Thomas construction, become obvious in this decomposition. The form factors cannot be chosen such that they are only functions of the squared 4-momentum transfer at the $W$-meson vertex, they also depend on Mandelstam $s$, i.e. the invariant mass squared of the whole neutrino-$B$-meson (or equivalently electron-$D$-meson) system. This does not spoil Poincar\'e invariance of the scattering amplitude, it just means that the $WBD$-vertex is also influenced by the presence of the scattering lepton. The Mandelstam-$s$ dependence may also be interpreted as a frame dependence of the $W^\ast B\rightarrow D$ subprocess. We consider two extreme cases, namely the minimum value of $s$ necessary to reach a particular $q^2<0$ and $s\rightarrow \infty$. The first choice corresponds to the Breit frame (BF), the second to the infinite-momentum frame (IF).

Using simple harmonic-oscillator wave functions with the oscillator parameters and masses taken from a front-form calculation of weak $B\rightarrow D$ transition form factors in the time-like region~\cite{Cheng:1996if} we obtain the results that are plotted in Fig.~\ref{bdspacelike}. Whereas the differences between IF and BF are small for $F_1$, they can be sizeable for $F_0$. From a physical point of view the IF seems to be preferable: unlike the BF, so called \lq\lq Z-graphs\rq\rq\ are suppressed in the IF and it is thus not necessary to take them into account, the Mandelstam-$s$ dependence vanishes for $s\gg m_B^2$, and our IF results agree with front-form calculations in the $q^+=0$ frame.
\vspace{-0.2cm}
\subsection{Time-like momentum transfers}
\label{sec:fftime}
\begin{figure}[t!]
\centering
\includegraphics[width=0.42\textwidth]{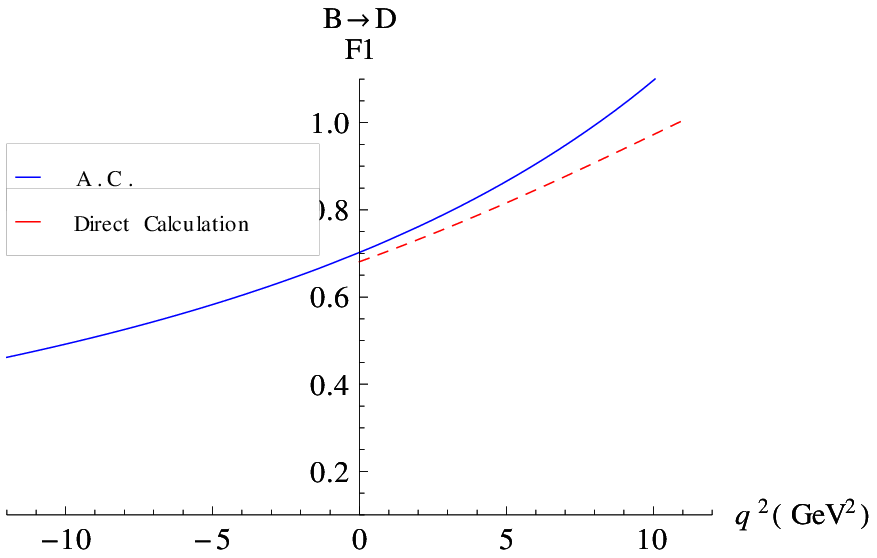}\hfill
\includegraphics[width=0.42\textwidth]{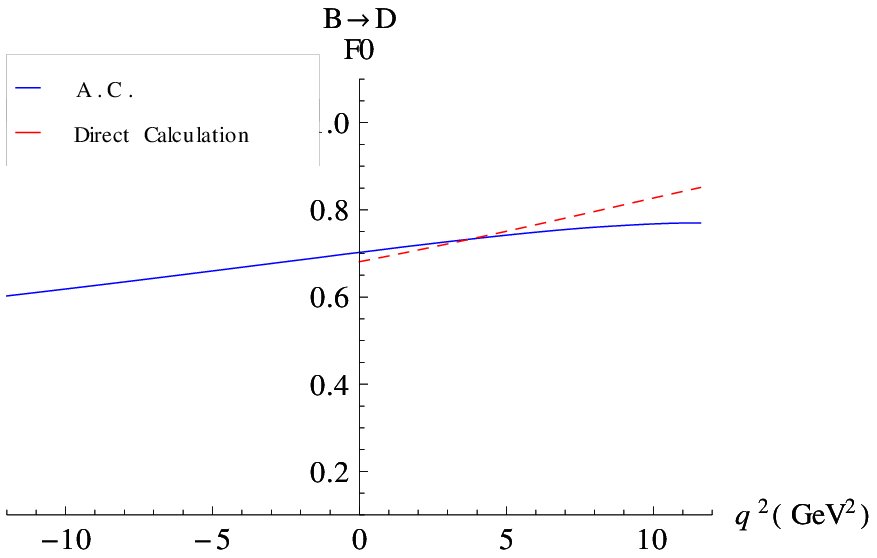}
\caption{The weak form factors $F1$ and $F0$ for the $B\rightarrow
D$ transition for  space- and time-like momentum transfers. The
solid line refers to the analytic continuation from space- to
time-like momentum transfers by making the replacement
$Q\rightarrow iQ$ in the infinite-momentum-frame result. The
dashed line is the outcome of a decay calculation in the $B$ rest
frame~\cite{GomezRocha:2012zd,Gomez-Rocha:2013bga}.} \label{bdtime}
\end{figure}
In the time-like momentum transfer region these form factors can be measured in semileptonic weak decay processes. Theoretically it is straightforward to adapt our relativistic multichannel approach such that one can deal with decay processes like $B\rightarrow D e \bar\nu_e$. Working in the velocity-state representation the decaying $B$-meson has to be at rest. For this kinematical situation it is, however, known from front-form calculations~\cite{Choi:1999bg} that non-valence contributions leading to  Z-graphs may become important. As already mentioned, a similar observation can be made for space-like momentum transfers, if the form factors are calculated in the BF and not in the IF (see also Ref.~\cite{Simula:2002vm}). In order to avoid problems with Z-graphs it is thus suggestive to take the form factor expressions obtained in the IF for space-like momentum transfers and continue them analytically to time-like momentum transfers by the replacement $Q\rightarrow iQ$. This is done in Fig.~\ref{bdtime} (solid lines). In this figure the results of a direct decay calculation~\cite{GomezRocha:2012zd,Gomez-Rocha:2013bga} (Z-graphs absent) are shown for comparison (dashed lines). The differences may be considered as an estimate of the size of Z-graph contributions. From the considerations just made it is clear that the solid line should be closer to experiment than the dashed one. This is indeed the case. What one knows experimentally is the slope of $F_1$ at zero recoil as measured in $B\rightarrow D e \bar\nu_e$ decays~\cite{Asner:2010qj}. It agrees approximately with the value which we get from our analytic continuation, whereas the decay calculation provides a much smaller value.

\vspace{-0.2cm}
\section{Outlook}
\label{sec:outlook}
\begin{figure}[t!]
\centering
\includegraphics[width=0.45\textwidth]{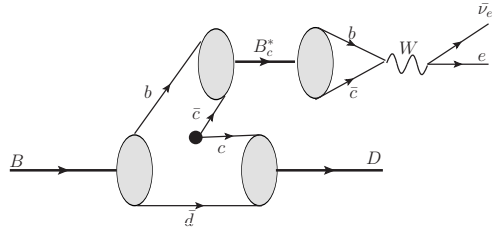}
\caption{Z-graph contribution to the 1$W$-exchange amplitude of
the  semileptonic $B\rightarrow D e \bar\nu_e$ decay}
\label{zgraph}
\end{figure}

Having estimated the size of Z-graph contributions indirectly, our
next task is now the explicit inclusion of Z-graphs in the decay
calculation.  One possible time ordering of a Z-graph contribution
to the $B\rightarrow D \, e\, \bar\nu_e$ decay is shown in
Fig.~\ref{zgraph}. It obviously requires the creation of a
$c\bar{c}$-pair, which we plan to describe by means of a $^{3}P_0$
model~\cite{Segovia:2012cd}. Within our relativistic
coupled-channel approach such a pair creation is easily
accommodated by an additional channel and a vertex interaction
obtained from the field-theoretical $^{3}P_0$ interaction
Lagrangean~\cite{Klink:2000pp}. Assuming instantaneous confinement
between the respective quark-antiquark pairs (indicated by blobs
in Fig.~\ref{zgraph}) on ends up with a simple,
vector-meson-dominance-like physical picture. The $Z$-graph
contribution to $B\rightarrow D$ decay can be understood as a
$B\rightarrow D$ transition induced by the emission of a
$B_c^\ast$ and, less important, excited ($b\bar{c}$) vector
mesons, which subsequently decay into $e\bar{\nu}_e$ by means of a
$W$. This simplifies the calculation of Z-graphs considerably,
since the main part can be done on the hadronic level. Only the
$B_c^\ast B D$-vertex (i.e. coupling strength and form factors) and the $B_c^\ast$ decay
has to be determined on the quark level. Proceeding in this way we
hope to achieve a more quantitative estimate of Z-graph
contributions in weak meson decay form factors.

\vspace{-0.2cm}
\begin{acknowledgements}
M. G\'omez-Rocha acknowledges the support of the \lq\lq Fonds zur
F\"orderung der wissenschaftlichen Forschung in \"Osterreich (FWF
DK W1203-N16 and in part under project P25121-N27).\end{acknowledgements}

\end{document}